\documentstyle[preprint, aps]{revtex}
\begin{document}
\draft
\title{ Effects of nuclear spins on the coherent evolution of a phase qubit. }
\author{ Geordie Rose and Anatoly Yu. Smirnov }
\address{ D-Wave Systems Inc. 320-1985 West Broadway,\\
 Vancouver, B.C. V6J 4Y3, Canada }
\date{\today}
\maketitle
 
\begin{abstract}
{
The role of nuclear spins in decoherence and dephasing of a solid state phase qubit is investigated. 
Both effects of  static spin environment and  spin polarization fluctuations in time are considered 
on the basis of non-Markovian Langevin-Bloch equations. 
We find conditions when coupling of a phase qubit to a bath of nuclear spins 
does not impair  coherent evolution of the qubit. }
\end{abstract}


\narrowtext

{\bf I. INTRODUCTION }

An analysis of decoherence and dephasing in quantum systems interacting with an environment has attracted a great attention in the recent years. This problem is of particular importance for a physical implementation of quantum computers, and, especially, for solid-state designs of qubits \cite{1,2}. Electronic or nuclear spins embedded into solid state materials
(see papers by D.G. Cory {\it et.al.}, G. Burkard {\it et.al.}, A. Imamoglu, and by  B.E. Kane in Ref.\cite{1})
provide a natural way for a practical realization of quantum bits. However, difficulties associated with a manipulation 
of individual spins render the spin-based design quite challenging. Superconducting qubits \cite{2,3,4,5,6,7} are more controllable, besides that they can be produced by means of well-developed litographic methods and offer advantages of scalability and flexibility. Unfortunately, operating states of a superconducting charge qubit extensively studied in the last time \cite{2,3} are highly coupled to environment degrees of freedom that leads to the high level of decoherence in the system. To get around this obstacle several models of phase superconducting qubits were proposed \cite{4,5,6,7}. These qubits utilizing superconducting phase as a dynamical variable with two macroscopically distinguishable values are effectively decoupled from charge degrees of freedom. 
In the general case the characteristic  property of the phase qubit is the presence of a small magnetic flux which has two opposite directions in the different states of the qubit. An interaction of the magnetic field created by the flux with a bath of nuclear spins will be an important mechanism resulting in destroying a quantum superposition of qubit's states and in statistical dispersion of characteristics of different qubits. 
In this work we develop a mathematical model describing the contribution of nuclear spins to  decoherence and inhomogeneous broadening of phase qubits and determine conditions when nuclear spins have only a slight effect on the operation of the qubit. The problem of the spin-bath mediated decoherence in phase qubits has been previously discussed in Refs.\cite{8,9,10}. 
In papers \cite{8,9} particular attention has been given to the case of static spin environment 
when the nuclear or paramagnetic spins have been subjected to the action of many magnetic fields, including the 
magnetic field from the qubit together with an external magnetic field and dipolar fields from other spins. 
By means of instanton technique Prokof'ev and Stamp \cite{9} have obtained, in particular, some estimations for the dephasing rates of the phase qubit in the static spin environment. It should be noted that the simple model considered in our paper allows us not only to get the same estimations but also to describe in detail a time evolution of the qubit's variables, and, especially, to determine the spectrum of all possible frequencies of quantum beats in the ensemble of qubits. This is a must for a correct analysis of logical operations in the ensemble of qubits.  
We suppose here that the magnetic field created by the qubit dominates over other magnetic fields acting on the nuclear spins. This assumption would hold for the qubit's design based on small grain-boundary junctions between d-wave superconductors \cite{5,6,11,12}.  
 
Besides the static case, in the present paper we consider a contribution of spin-lattice relaxation to the decoherence rate of the phase qubit. This phenomenon which is completely ignored in the above-mentioned paper \cite{9} leads to true decoherence \cite{13} as distinct from dephasing caused by the static spins. A qualitative estimation of effects 
of time-dependent spin fluctuations on the superconducting persistent current qubit has been performed in Ref.\cite{10}. 
In the paper \cite{10} nuclear spins are supposed to be in the state of thermal equilibrium despite the fact that the qubit's magnetic field oscillates with a high frequency determined by splitting of two lowest energy states. Under such conditions an application of the fluctuation-dissipation theorem for the calculation of the qubit's decoherence rate as it has been done in Ref.\cite{10} casts some doubt. A longitudinal relaxation of nuclear spins takes long enough time $T_1$, however, many nuclear spins will contribute to qubit's decoherence, so that the total decoherence rate of the qubit will be proportional not only to the damping rate of a single nuclear spin $T_1^{-1}$ but also to the number of spins in the qubit's area. It follows herefrom that the mechanism of qubit's decoherence due to spin-lattice relaxation is competitive with other mechanisms and, at least, can not be rejected without more circumstantial consideration. 

The present paper is organized as follows.
The next section is devoted to the formulation of the problem. In Sec. III we analyze inhomogeneous broadening in the ensemble of qubits caused by a static spin environment.
Decoherence rates of the qubit interacting with a spin polarization fluctuating in time are calculated in Sec. IV.  
\\

{\bf II. MODEL } 

As mentioned above, there is a spontaneous flux 
$ (\Phi \sim 10^{-2} - 10^{-3} $  $\Phi_0 )$ concentrated in the central 
part of the phase qubit and having opposite directions in two 
degenerate equilibrium states $\pm \varphi_0 $ \cite{4,5,6,7}. 
This flux creates the corresponding magnetic field ${\bf B}_0({\bf r})$ which
also  has two directions: along and opposite of z-axis, $\pm B_0({\bf r})$ at
different values  of the qubit's phase. 
The existence of two-well potential for the phase difference in the asymmetric grain boundary Josephson junctions 
in d-wave cuprates directly follows from the experimental results obtained in Ref.\cite{11}.
A doubly-degenerate ground state of such a junction is split into the symmetric and antisymmetric states with 
$\Delta $ being the energy difference between them. 
The qubit's two-level system is conveniently described by the Pauli's matrices 
$\tau_x, \tau_y, \tau_z, $ so that the tunneling Hamiltonian 
is given by the expression: $ H_0 = (\Delta /2) \tau_z. $ 
The eigenvalues $\pm 1 $ of the matrix $\tau_x$ corresponds to the positioning of the qubit in the right well $(\varphi_0)$ 
or in the left potential well $(-\varphi_0),$ whereas the eigenvalues of the matrix $\tau_z$ corresponds the energy levels 
$\pm (\Delta /2 )$ of the antibonding (antisymmetric) and bonding (symmetric) states.  
The magnetic field produced  by
the qubit can be also represented as a matrix: $ \hat{B}_0 = \tau_x B_0({\bf
r})$ with $\int B({\bf r}) dS = \Phi .$ 

Nuclear spins in the area of the qubit are subjected to the action of this magnetic field, 
and their energy operator looks as follows \cite{10} 
\begin{equation}
H_s = - \tau_x \sum_i g \mu_{Ni} B(r_i) (\sigma_i)_z
\end{equation}
where $\mu_{Ni} $ is the nuclear magneton, 
$(\sigma_i)_z$ is z-projection of the Pauli's matrix describing $i-th$ nuclear
spin, $B(r_i) $ is the magnetic field induced by the qubit in the point $r_i$
where $i-th$ nuclear spin is located. Here we take a sum over 
all nuclear spins placed in the qubit's area. The effect of any external magnetic field on the nuclear spins is supposed to be small compared to the action of the qubit's magnetic field. 
It might be well to point out that the Hamiltonian (1) can describe an interaction of the qubit not only with nuclear spins but with any localized magnetic moments as well, and, in particular, with an electronic spin 1/2 degree of freedom located in the $CuO_2$ planes of the YBCO superconductor \cite{14}. In this case we have to extend the sum in Eq.(1) to the electronic spins with the substitution of electronic magnetons $\mu_{Bi}$ for $\mu_{Ni},$ and the consideration $\sigma_i$ as electronic Pauli's matrices.  

To study an internal dynamics and fluctuations in the system of nuclear
spins we suppose that $i-th$ nuclear spin is coupled to the dissipative
environment with  variables $\{Q_x(t), Q_y(t), Q_z(t)\}_i $. This coupling has
the form 
\begin{equation}
H_{N} = - \sum_i {\bf \sigma }_i \cdot {\bf Q}_i.
\end{equation}
For example, a heat bath of acoustic phonons is described by the variable \cite{15}
$$
{\bf Q }_i = V^{-1/2} {\bf G} \sum_q \sqrt{\hbar |{\bf q}|\over \rho c_s} i (b_{\bf q} - b_{\bf -q}^+)e^{i{\bf q\cdot r_i}}
$$
Here $b_{\bf q}^+, b_{\bf q}$ are creation-annihilation operators of acoustic phonons with 
wavevectors ${\bf q}$ and the sound velocity $c_s;$ ${\bf G}$ is a constant of spin-phonon coupling, 
$\rho$ and $V$ are the density and the volume of the crystal, respectively. The spin-phonon interaction (2), while unavoidable, is weak enough, especially for the nuclear spins having a small magnetic moment. As a result, the directions of nuclear spins do not conserve exactly. But due to the weakness of spin-phonon coupling we can suppose in the first approximation that nuclear spins have fixed directions which, however, vary from one qubit to another. 
In the part III we investigate inhomogeneous broadening or dephasing in the ensemble of qubits associated with this variation. Time-dependent spin fluctuations due to the spin-phonon interaction lead to the true decoherence of the qubit. 
The section IV is devoted to the study of this phenomenon. It is of interest that the spin-lattice relaxation for the copper, oxygen and yttrium nuclei is usually considered as a result of the interaction between nuclear spin and electronic spins localized in the $CuO_2$ planes\cite{14}. Besides that, the electronic spins contribute directly to the decoherence of the qubit with the polarization, $P,$ containing the time-dependent part only.  

The total Hamiltonian of our system will look like:
\begin{equation}
H = {\Delta \over 2} \tau_z - P \tau_x - 
\sum_i {\bf \sigma}_i \cdot {\bf Q}_i + H_B.
\end{equation}
Here 
\begin{equation}
P = \sum_i {\Delta_i \over 2} (\sigma_i)_z
\end{equation}
is the operator of Zeeman energy of nuclear spins in the fixed qubit's magnetic field, 
$\Delta_i = 2 g \mu_{Ni} B(r_i)$ is the Zeeman frequency of $i-$th nucleus,
and $H_B$ is the Hamiltonian of the free heat bath. 
In the case of acoustic phonons the Hamiltonian $H_B$ will look like 
$$
H_B = \sum_q \hbar c_s |{\bf q}| (b_{\bf q}^+ b_{\bf q} + 1/2)
$$ 
We suppose that the free heat bath (without the interaction with the spins) is in thermal equilibrium with 
a temperature $T.$ The operator $P$ (4) is proportional to the spin polarization in z-direction. Because of this, we will mention this operator later just as a "spin polarization".

It follows from the Hamiltonian (3) that the time evolution of the qubit's quasispin described by the matrices ${\tau_x, \tau_y, \tau_z}$ is governed by the equations
\begin{eqnarray}
\dot{\tau}_x = - \Delta \tau_y, \nonumber\\
\dot{\tau}_y = \Delta \tau_x + 2 \tau_z P, \nonumber\\
\dot{\tau}_z = - 2 \tau_y P.
\end{eqnarray}

Direct and indirect interactions between different nuclear spins are neglected in this model.
In view of a great number of nuclear spins in the qubit's area an influence
of these spins on the qubit's dephasing can be quite considerable.
Fortunately, the main part of the spin polarization, $P_0$, is almost static
and determined by the initial spin configuration of the given qubit.   
This part differs in different qubits, as well as in the same qubit but
taken  at the time of the next logical operation. These fluctuations result in  statistical dispersion of properties of qubits followed by the impact on the entanglement of qubits. In the next section we will analyze a contribution of the static spin polarization without resorting to the perturbation theory.  
Besides the static part the
spin polarization will include a small part fluctuating in time. 
This part $\tilde{P}(t) = P(t) - P_0 $  is supposed to be here due to the spin-lattice 
interaction and results in destroying of quantum coherence of each individual qubit.  \\

\begin{center}
{\bf III. INHOMOGENEOUS BROADENING }
\end{center}

In this section we consider effects of the {\it static } spin
environment ($P = P_0$) on the time evolution  of the qubit's ensemble
without taking into account the component fluctuating in time. The interaction between
the qubit degrees  of freedom and the static nuclear spins is not supposed to
be weak. 
Each of the qubits has a specific distribution of nuclear spins. We have to average qubit's
operators over all  configurations of nuclear spins to obtain an information about a time evolution 
of the qubit's ensemble, for example, an information about frequencies of quantum
beats (tunneling oscillations) of different qubits. Our goal is to find conditions when the frequency dispersion of qubit's ensemble due to the interaction with nuclear spins  will not influence on a performance
of logical operations, especially, on the entanglement  of different qubits. 

In this case the time evolution of the qubit's operators is described by the equations (5)
with the time-independent spin polarization $P = P_0.$
A direct solution of the system (5) of differential equations (with $P=P_0$)    
gives us the formula 
\begin{eqnarray}
\tau_x(t) = \left[ 1 - {\Delta^2 \over \Omega^2} 
( 1 - \cos \Omega t ) \right]\tau_x(0)  - \nonumber\\
\Delta {\sin \Omega t \over \Omega }\tau_y(0)  - {2 \Delta P_0 \over \Omega^2 }
(1 - \cos \Omega t) \tau_z(0)
\end{eqnarray}
that represents oscillations of qubit variables with a frequency  $\Omega^2 = \Delta^2 + 4 P_0^2$ in a specific spin environment characterized by the parameter $P_0.$ 
However, this parameter is changed when we proceed to the next qubit. To study a variety of possible evolutions we have to average 
the expression (6) over an ensemble of all spin configurations.
It has been done in the Appendix A.
As a result, the time evolution of the qubit's variable averaged over nuclear spin configurations is given by the following expression
\begin{eqnarray}
\langle\tau_x(t)\rangle  = 2^{-2N}  \sum_{m = -N}^{ m = N} C_{2N}^{N-m} 
\left[ 1 - \Delta^2 {1 - \cos(t\sqrt{4\Delta_0^2 m^2 + \Delta^2}) \over 
4\Delta_0^2 m^2 + \Delta^2} \right]\langle \tau_x(0)\rangle  - \nonumber\\ 2^{-2N} \sum_{m =
-N}^{ m = N} C_{2N}^{N-m}  \Delta {\sin(t\sqrt{ 4\Delta_0^2 m^2 + \Delta^2})
\over \sqrt{ 4\Delta_0^2 m^2 + \Delta^2}} \langle \tau_y(0)\rangle.
\end{eqnarray}

In the case of zero Zeeman frequency $(\Delta_0 = 0)$ the phase differences of all identical qubits 
in our ensemble will oscillate with the same frequencies equal to tunnel splitting $\Delta.$ 
Otherwise, the frequencies of quantum beats 
of different qubits can take the spectrum of values 
$\omega_m = \sqrt{ 4\Delta_0^2 m^2 + \Delta^2}$ with $ m = 0,1,.., N,$ because each of these qubits has 
the own quasistatic distribution of nuclear spins (up or down), and, therefore, the own energy of interaction with these spins. Changing an orientation of a single nuclear spin, e.g., from up to down, will change the interaction energy between this spin and the qubit by the value of $\Delta_0.$ This fact results in the discrete character of qubit's frequencies.   
During the time of one logical operation the frequency of the given 
qubit is equal to one of the frequencies $\omega_m.$ 
But in the time of performing of the next operation separated from 
the first operation by the interval that is in excess of the spin-lattice 
relaxation time $T_1$  this qubit can be characterized by another 
frequency from the set $\{\omega_m \}.$ 
The adjacent qubits can also have different frequencies that will affect entanglement processes.  
The dispersion of qubit's freqiencies is described by the binomial coefficients 
$$C_{2N}^{N-m} = {(2N)! \over (N - m)!(N + m)! } $$
times $2^{-2N}.$ At large $N$ and $ m \ll N $ this function is approximated
by the Gaussian distribution:
$$
2^{-2N} C_{2N}^{N-m}  \simeq \exp( - m^2 /N)
$$
so that the probable values of $m$ are in the range from zero to $\sqrt{N}.$ 
The maximum frequency of quantum beats is of order  
$\sqrt{ 4\Delta_0^2 N + \Delta^2}$. 
In the practically interesting case of weak coupling between nuclear spins and the qubit we obtain for the inhomogeneous 
broadening of the qubit's line
\begin{equation}
(\Delta \omega)_{weak} = 2N {\Delta^2_0 \over \Delta },
\end{equation}
whereas for the strong coupling we have 
\begin{equation} 
(\Delta \omega)_{strong} = {\Delta^2 \over \Delta_0 \sqrt{2N} }. 
\end{equation}
In the framework of instanon technique Prokof'ev and Stamp \cite{9} have considered more general model of qubit's dephasing due to static nuclear spins, but at the final stage they have obtained the same estimations as ones in Eq.(8),(9). 
This points to the ability of our simple mathematical model to describe properly the above-mentioned mechanism of qubit's dephasing. It should be emphasized that besides of estimations our model gives the mathematical expression for the time evolution of the qubit's variable $\langle \tau_x (t) \rangle $ (7) that is of great importance for an analysis of logical operations in the ensemble of qubits.

The static nuclear spin environment does not affect 
the time evolution of the qubit and, therefore, the performance of logical operations, 
if the number of nuclear spins, $2N$, interacting with the qubit is less than the ratio of 
 tunnel splitting to Zeemann splitting squared:
\begin{equation}
2N < {\Delta^2 \over 2\Delta_0^2}.
\end{equation} 

For the sake of estimation we consider the qubit's model with the linear size of the central area  
$l \simeq 10^{-5} cm. $ This size seems to be too small for designs considered in \cite{4,7}, but 
it is very reasonable for qubits based on Josephson junctions between d-wave superconductors \cite{5,6,11,12}. 
As shown in \cite{12}  for this design the magnetic field produced by circulating spontaneous currents can be 
localized in the area with one size of about $10 \xi_0$ ($\xi_0$ is the coherence length, $\xi_0 \sim 10^{-6} cm$) 
and another size of about the width of the junction $(\sim 0.2 \mu m)$. Such small junctions are necessary also 
to eliminate experimentally defects and other irregularities.    
Even with both sizes of order $0.2 \mu m$   
and with the magnetic flux created by the qubit of order $\Phi \simeq 10^{-2} \Phi_0, \Phi_0 = hc/2e$ we obtain that the qubit's magnetic field can reach the value $10$ $ G$ comparable to the dipole fields between the nuclear spins. 
Different components of $YBa_2Cu_3O_7$ have magnetic moments in the range from $-0.238 (^{89}Y)$ to $3.075 (^{65}Cu)$ of nuclear magneton. Because of this, for the Zeeman frequency of one
nucleus  in the qubit's magnetic field we get the estimation $\Delta_i \simeq 10^4 - 10^5
(1/s), $ so  $\Delta_i / \Delta \simeq 10^{-3} - 10^{-4} \ll 1,$ if $\Delta \simeq 10^8
(1/s).$ As follows from Eq.(10), the number of nuclear spins 
coupled to the qubit should be less than $10^6 - 10^7$ if we want to get
rid of effects of above-mentioned inhomogeneous broadening on the qubit's
evolution. The magnetic field of the qubit can not penetrate into the superconductors at the depth in excess of the penetration length $\lambda$. Only nuclear spins located inside of the volume with third dimension smaller than the penetration length will be subjected to the action of the qubit's magnetic field. 
For a $YBa_2Cu_3O_7$ sample with cell sizes of order $(3, 3, 11) \AA $ 
   there is almost $2N \sim 10^8 $ nuclear spins in the qubit 
if the penetration length $\lambda \sim 10^{-5} cm.$ This number of nuclear spins is 
close enough to the above-mentioned requirements. With small decreasing the number  of spins 
(or decreasing the ratio $\Delta_0^2 / \Delta^2$)  
the statistical dispersion of frequencies of different qubits will be of little importance and the operation of the
qubit under discussion will not be distorted by this mechanism
of inhomogeneous broadening. It should be mentioned that the qubit's magnetic field and, therefore,  
the Zeeman frequency of a single nucleus decreases with increasing the size $l$ of the qubit: $\Delta_0 \sim \Phi /l^2,$
whereas the number of nuclear spins, $N,$ is proportional to the qubit's area, $l^2$: $N \sim l^2\times \lambda, $ provided the qubit's magnetic flux, $\Phi,$ and the penetration length, $\lambda,$ remain constant. Then, the inhomogeneous broadening of the line (8) is inversely proportional to the area of the qubit: $(\Delta \omega)_{weak} \sim 1/l^2,$ and we can expect that for larger size qubits studied in Refs. \cite{7,10} the effects of nuclear spins are of lesser importance. However, in this case the magnetic field created by the qubit can be less than the dipole field between the nuclear spins, and we need the generalization of the formulae (8),(9), (10) to describe the situation correctly.   \\

\begin{center}
{\bf IV. DECOHERENCE OF PHASE QUBIT IN THE SPIN BATH}
\end{center}

Now we analyze an effect of a nuclear spin polarization fluctuating in time on
decoherence processes in the superconducting qubit. The contribution of electronic spins localized in $CuO_2$ planes of the YBCO superconductor \cite{14} to the qubit's decoherence rate can be calculated by the same approach. 
We assume here that the concentration of nuclear spins 
are sufficiently small, so that the condition (10) is fulfilled and we can neglect an effect of static spin polarization on the evolution of the qubit: $P_0 = 0,$ so that fluctuations of the spin polarizations are completely due to coupling of nuclear spins to the heat bath, e.g. to phonons.  
Now a contribution of the
fluctuating component to the total Hamiltonian (3) of the system is supposed to be small compared to the tunneling term
$(\Delta/2)\tau_z.$
Here we consider also weak coupling of nuclear spins to a dissipative
environment, for example, to a phonon heat bath, as a source of fluctuations of nuclear spin polarization. 
This coupling results in the
energy dissipation and thermalization of the spin system. In the weak-damping
approximation when the energy of the interaction
between spins and the heat bath is much less than the Zeeman frequency
$\Delta_i$ we can find a dissipative evolution of nuclear spin operators ${\sigma_z}_i$  in the flipping magnetic field $\tau_x B_z(r_i)$ created by
the qubit and calculate a correlation function 
$$\langle
[\tilde{P}(t),\tilde{P}(t')]_{\pm}\rangle = \langle \tilde{P}(t)\tilde{P}(t') \pm \tilde{P}(t')\tilde{P}(t)\rangle $$ with the polarization $\tilde{P}(t)$ being determined by Eq.(4):
$$ 
\tilde{P}(t) = \sum_i {\Delta_i \over 2} [(\sigma_i)_z - \langle (\sigma_i)_z \rangle ].
$$ 
Here brackets $\langle ..\rangle $ mean averaging over the state of the equilibrium heat bath.  
It should be noted that the equations (5) descriptive of the qubit's evolution incorporate the total spin polarization 
$P = \tilde{P} + \langle P\rangle. $ However, the mean part of the polarization $\langle P \rangle $ leads only to small 
corrections for the tunneling frequency $\Delta,$ and, therefore, it can be dropped out.  
At the same time the fluctuating part of the spin polarization $\tilde{P}$ causes the destruction of the qubit's coherence 
for the time interval which is equal to the relaxation time of the
averaged $x-$projection $\langle \tau_x\rangle $ of the qubit's quasispin.
The evolution of matrices ${\tau_x, \tau_y, \tau_z }$ is governed by the equations (5) where $P = \tilde{P}(t).$
The fluctuating part of spin operator $\tilde{P}(t)$ includes contributions from many different spins
and can be described approximately by Gaussian statistics.
With this property in mind  the mean value of the product of the qubit's matrix and the spin polarization, say, 
$\langle \tau_z \tilde{P}\rangle, $ can be expressed as \cite{16,17,18}
$$
\langle \tau_z \tilde{P}\rangle = \langle {1\over 2}[\tau_z(t),\tilde{P}(t)]_+\rangle = 
\int dt_1\tilde{M}_P(t,t_1)\langle i [\tilde{\tau}_z(t), \tilde{\tau}_x(t_1)]_-\rangle .
$$
Here 
\begin{equation}
M_P(t,t_1) = \langle{1\over 2}[\tilde{P}(t),\tilde{P}(t_1)]_+\rangle 
\end{equation} 
is the correlation function of the spin bath in the presence of flipping magnetic field created by the qubit, 
$[A,B]_{\pm} = AB \pm BA,   \tilde{M}_P(t,t_1) = M_P(t,t_1)\theta (t - t_1)$ with $\theta (t - t_1) $ 
being the Heaviside step function.  
Averaging Eq.(5) over fluctuations of the Gaussian spin variable
$\tilde{P}(t),$ i.e. eventually over the heat bath fluctuations, gives 
the non-Markovian equations \cite{16,17,18}  for the mean qubit's variables
\begin{eqnarray}
\langle \dot{\tau}_x\rangle  = - \Delta \langle \tau_y \rangle , \nonumber\\
\langle \dot{\tau}_y\rangle  = \Delta \langle \tau_x \rangle + 2 \int
dt_1\tilde{M}_P(t,t_1)\langle i [\tau_z(t), \tau_x(t_1)]_-\rangle , 
\nonumber\\
\dot{\tau}_z = -2 \int
dt_1\tilde{M}_P(t,t_1)\langle i [\tau_y(t), \tau_x(t_1)]_-\rangle .
\end{eqnarray}
As mentioned above time-dependent fluctuations of spin polarizations are supposed to be weak. 
Because of this, to calculate the commutators in Eqs.(12) we can use the equations 
$$
\ddot{\tau}_x + \Delta^2 \tau_x = 0  
$$
for the free oscillations 
of qubit's variables without any spin polarization. 
Then, the commutators are expressed in terms of qubit's matrices, for example, 
$$ 
i [\tau_z(t), \tau_x(t_1)]_ = -2 \tau_y(t) \cos \Delta(t-t_1) + 2 \tau_x(t) \sin \Delta(t-t_1). 
$$
Ignoring the frequency shift of the qubit due to spin fluctuations we obtain the simple equation 
\begin{equation}
\langle \ddot{\tau}_x\rangle  + 2\Gamma \langle \dot{\tau}_x\rangle  + \Delta^2 \langle\tau_x\rangle = 0  
\end{equation} 
describing the relaxation of qubit's oscillations between two wells with the relaxation rate 
\begin{equation}
\Gamma =  S_P(\Delta ),
\end{equation}
where 
\begin{equation}
S_P(\omega) = \int_{-\infty}^{\infty} d\tau e^{i\omega \tau } M_P(\tau ) 
\end{equation}
is the spectral density of spin fluctuations. 
The relaxation of the qubit matrix $\tau_x $ to the equilibrium value is
described by the expression :
\begin{equation}
\langle \tau_x(t)\rangle = \left[ \langle \tau_x(0)\rangle \cos(\Delta t) - 
\langle \tau_y(0)\rangle  \sin(\Delta t)\right] e^{-\Gamma t}. 
\end{equation}

The coefficient $\Gamma $  (14) represents also the rate of qubit's decoherence in the spin
environment. 
To find this decoherence rate we have to calculate the spectral function $S_P(\omega)$ (15) and the correlation function 
$M_P(t,t_1)$ (11). Fluctuations of different spins are supposed to be independent. In view of this fact together 
with Eq.(4) the correlator $M_P(t,t_1)$ is proportional to the sum of correlation functions of $z$-projections of nuclear spins:
\begin{equation}
M_P(t,t_1) = \langle {1\over 2}[\tilde{P}(t),\tilde{P}(t_1)]_+\rangle =
\sum_i {\Delta_i^2 \over 4}\langle {1\over 2} [(\sigma_i)_z(t),(\sigma_i)_z(t_1)]_+\rangle . 
\end{equation}
Note that only $z-$projection of the nuclear spins is coupled to the qubit because the qubit's magnetic field is
always directed  parallel to z-axis while its magnitude oscillates with time.  
The calculation of the correlation function of $i$-th spin 
$ \langle (1/2) [(\sigma_i)_z(t),(\sigma_i)_z(t_1)]_+\rangle $ in the presence of flipping magnetic field and a heat bath together with finding the spectral density of spin fluctuations $S_P(\omega )$ (15) has been performed in the Appendix B.

With the formulas (14),(B21) we obtain the following expression for the decoherence rate (or the inverse decoherence time $\tau_d^{-1}$ ) of the phase qubit 
\begin{equation}
\Gamma = {1 \over \tau_d} =  {1\over 2 } \sum_i \Delta_i^2 {\gamma_i(\Delta ) \over \Delta^2
+ \gamma_i^2(\Delta ) }.
\end{equation}

To obtain the expression (18) we have supposed that the decoherence rate, $\Gamma,$ is much less than the qubit's tunneling frequency, $\Delta,$ and that the the linewidth of i-th nuclear spin, $\gamma_i,$ is less than the corresponding frequency $\Delta_i$: $\Gamma \ll \Delta, \gamma_i \ll \Delta_i.$ 

In view of the fact that the relaxation rate of i-th nuclear spin,
$\gamma_i(\Delta ), $ or, in other words, an inverse longitudinal relaxation time $(1/T_1)_i$ of $i$-th spin, 
 is much less than the qubit's frequency $\Delta:  (1/T_1)_i = \gamma_i(\Delta ) \ll \Delta , $ we find that the
qubit's decoherence rate is proportional to the sum of spin line widths
$\gamma_i$ due to spin coupling to a heat bath taken with the weights 
$(\Delta_i^2 /\Delta^2 ):$ 
\begin{equation}
{1 \over \tau_d} =  {1\over 2}\sum_i {\Delta_i^2
\over \Delta^2 } \gamma_i(\Delta ) = \sum_i {\Delta_i^2 \over \Delta^2 } 
\sum_m J_m\left( 2 {\Delta_i \over \Delta }\right) 
\left[ S\left(\Delta - {m\Delta \over 2} \right) + S\left(\Delta + {m\Delta
\over 2} \right)\right].
\end{equation} 
This formula takes into consideration
oscillations of the qubit's magnetic field with the tunneling frequency
$\Delta .$ Usually Zeeman splitting $\Delta_i$ of i-th nuclear spin in the
qubit's magnetic field as well as in the magnetic field created by other nuclear spins 
 is much less than the frequency of quantum beats
$\Delta : \Delta_i \ll \Delta .$ Because of this, the term with $m=0$ gives
the major contribution to the sum in Eq.(19), and in this approximation the
qubit's decoherence rate will look like:
\begin{equation}
{1 \over \tau_d} =  2\chi''(\Delta ) \coth\left({\Delta \over 2T}\right) 
\sum_i {\Delta_i^2 \over \Delta^2 } .
\end{equation}
Here $ \chi''(\omega ) $ is the imaginary part of the susceptibility of the
heat bath with the temperature $T.$ The interaction of nuclear spins with this
heat bath, for example, with lattice vibrations, provides a mechanism of
dissipation and thermalization in the spin system. If the heat bath
temperature is in excess of the tunnel splitting: $T\gg \Delta,$ the qubit's
decoherence rate increases linearly with temperature:
\begin{equation}
{1 \over \tau_d} =  4T {\chi''(\Delta )\over \Delta } \sum_i
{\Delta_i^2 \over \Delta^2 } 
\end{equation}
The nuclear spin-lattice relaxation time is usually measured 
at the field of order $10^4 Gs$ when the Zeeman frequency for one of our nucleus 
is $10^7 (1/s).$
At this field and at the temperature $T \simeq 1 K$ 
the spin-lattice relaxation rate is of order or less than $10 (1/s)$ \cite{19}.
According to results by Morr and Wortis \cite{20} 
the spin-lattice relaxation rate is approximately proportional 
to the applied magnetic field squared. Our decoherence time, $\tau_d $ (19)-(21), 
is determined by the spin-lattice relaxation rate $\gamma_i(\Delta )$  taken at the frequency 
$\Delta $. For the qubit's model mentioned in the end of Sec.III  we have $\Delta \simeq 10^8 (1/s),$  
that  is more than the usual frequency $10^7 (1/s)$ of relaxation 
time measurements by a factor of 10. So, the spin-lattice relaxation rate
corresponding  to our quantum beats frequency $\Delta $ will be estimated as 
$(1/T_1)_i = \gamma_i(\Delta ) \sim 10^3(1/s).$
For the same $YBa_2Cu_3O_7$ sample as in Sec.III  with 
$2N \sim 10^8 $ nuclear spins the qubit's 
decoherence rate due to magnetic coupling to the nuclear spins is 
$$ {1 \over \tau_d } \simeq  10^3 (1/s). $$
This decoherence time is much more than the period of quantum beats oscillations:
$\tau_d \Delta \sim 10^5 \gg 1, $ so that the necessary condition 
$$ \tau_{tunneling} < \tau_{gate } < \tau_{decoherence }$$ 
or 
$$ 10^{-8} s < \tau_{gate } < 10^{-3} s , $$ 
holds if  the nuclear spins are considered as the main decoherence mechanism.
Nevertheless, as is evident from the foregoing, the mechanism of the qubit's decoherence related to the 
spin-lattice relaxation can not be considered as improbable, especially as this mechanism causes true decoherence of a single qubit in distinction to dephasing occuring in the ensemble of qubits. 
It is necessary to
emphasize that the role of localized magnetic moments related to the electronic shells in the
superconductor has to be investigated as well for the correct
estimation of the decoherence rate $\tau_d^{-1}.$ 
To estimate the contribution of electronic spins belonging to the  $Cu$ and $O$ ions to the decoherence of the qubit 
we can assume that the ratio between the electronic spin magnetic moment and the nuclear magneton is about $10^3,$ so that for parameters mentioned at the end of the part III, the Zeeman splitting of the electron in the qubit's magnetic field is of order $\Delta_i=\Delta_0 \sim 10^8 (1/s).$  The spin line width, $\gamma_i$ (B19), increases more significantly because it is proportional to the electronic magneton squared: $\gamma_i \sim 10^8-10^9 (1/s)$, and reaches the values of $\Delta_i$ and $\Delta.$ It is evident that for the correct analysis of this limit we have to generalize the formula (18), however, this task is beyond the scope of the our paper devoted to the effects of nuclear spins. Electronic spin magnetic moments contribute, in particular, to the spin-lattice relaxation of nuclear spins, and, by doing so, indirectly affects the decoherence rate of the qubit. It is of interest that the decoherence rate (18) grows with the value of the spin magnetic moment showing the tendency for a saturation. Because of this, we can expect that the localized electronic spins will not destroy the working conditions for the phase qubit. \\

\begin{center}
{\bf CONCLUSIONS.}
\end{center}

We have considered a contribution of nuclear spins to the dephasing and decoherence of a phase qubit. 
An effect of the static spin environment has been analyzed without resorting to the perturbation theory. 
For this case we have found  a condition (10)  when dephasing or inhomogeneous broadening in an ensemble of qubits 
has no effect on the operation of a quantum computer.  Decoherence rates (19),(20),(21) of the qubit weakly 
coupled to the heat bath of nuclear spins have been calculated as well; in so doing we have taken into account  
fluctuations of a spin polarization in time stemming from the interaction of nuclear spins with a basic dissipative environment, for example, with lattice vibrations. Interestingly, formulae for the relaxation rate $\gamma_i(\omega )$ (B19) and for a spectral density $K_i(\omega )$ of fluctuations of an individual nuclear spin in the time-varying magnetic field 
of the qubit have been derived in the process. These formulae are also useful for understanding the electron spin resonance 
in the flipping magnetic field. For the realistic parameters of the phase qubit based on the YBCO crystal \cite{5,6,12} we have estimated a decoherence time that meets the necessary condition for the successful operation of the qubit. \\

\begin{center}
{\bf ACKNOWLEDGEMENTS}
\end{center}
We would like to thank A.M. Zagoskin and M.H.S. Amin for many useful discussions. 

\appendix
\section{}
In the general case, averaging the expression (6) over an ensemble of all spin configurations, namely, finding the functions 
$\langle \sin \Omega t/\Omega \rangle$ and $\langle (1- \cos \Omega t)/\Omega^2 \rangle $,  can be done by means of Fourier transforms, when the dependence on $P_0$ is transferred from the frequency $\Omega^2 = \Delta^2 + 4 P_0^2$ to the exponent $ e^{i\xi P_0 },$ 
 for example, 
\begin{equation}
\langle {\sin \Omega t \over \Omega } \rangle = 
\langle \int d \xi e^{i\xi P_0 } F(\xi ) \rangle 
\end{equation}
with 
\begin{equation}
F(\xi ) = {1 \over 2 \pi} \int_0^{\infty} dx \cos(\xi x) { \sin 2t \sqrt{(\Delta /2)^2 + x^2}
\over \sqrt{(\Delta /2)^2 + x^2}}.
\end{equation}
After calculating the integral (A2), we get for $F(\xi )$ 
\begin{equation}
F(\xi ) = {1\over 4} J_0\left({\Delta \over 2} \sqrt{4t^2 - \xi^2 } \right) 
\theta (2t - |\xi | )
\end{equation}
where $\theta (2t - |\xi | ) $ is the Heaviside step function, and $J_0$ is the Bessel function. In the formula (A1) we average the function $\exp(i\xi P_0)$ or 
$\cos(\xi P_0)$ over the spin configurations:
\begin{equation}
\langle {\sin \Omega t \over \Omega } \rangle = 
{1 \over 2} \int_0^{2t} d \xi \langle \cos\xi P_0 \rangle J_0\left({\Delta \over 2} \sqrt{4t^2 - \xi^2 } \right).  
\end{equation}
The characteristic functional of the spin ensemble can be represented as 
\begin{eqnarray}
\langle e^{i\xi P_0}\rangle = 
\langle \exp\left( i \xi \sum_i {\Delta_i \over 2 }(s_i)_z \right)\rangle = \nonumber\\
\langle \exp\left( i {\Delta_1 \xi \over 2 }(s_1)_z \right) \cdot ...
\exp\left( i {\Delta_{2N} \xi \over 2 }(s_{2N})_z \right) \rangle = \nonumber\\
\cos\left({\Delta_1 \xi \over 2 }\right) \cdot ...\cos\left({\Delta_{2N} \xi \over 2 }\right)
 = \prod_{i=1}^{2N} \cos\left({\Delta_i \xi \over 2 }\right) .
\end{eqnarray}
Here $2N$ is the total number of the nuclear spins in the qubit's volume.
As a result, for the averaged function (A4) we obtain
\begin{equation}
\langle {\sin \Omega t \over \Omega } \rangle = 
{1 \over 2}  \int_0^{2t} d \xi  \cos\left({\Delta_1 \xi \over 2 }\right)  ...\cos\left({\Delta_{2N} \xi \over 2 }\right) 
 J_0\left({\Delta \over 2} \sqrt{4t^2 - \xi^2 } \right).  
\end{equation}
For the sake of simplicity we restrict ourselves to the case of the homogeneous magnetic field created by the flux: $B(r_1) = ... = B(r_{2N}),$ and $\Delta_1 = ... =\Delta_{2N} = \Delta_0.$ It should be mentioned that in the real case 
the dispersion in $\Delta_i$ can be of order of its average value $\langle \Delta_i \rangle .$ 
However, as it will follow from Eq.(10) derived later in the Section III, taking into account this dispersion will change an estimation for the 
dephasing rate $\Delta \omega $ by factor of order four that is negligible small compared to possible variation 
in the number of nuclear spins $2N \sim 10^7-10^{8}.$
With the above-mentioned approximation we get 
\begin{equation}
\prod_{i=1}^{2N} \cos\left({\Delta_i \xi \over 2 }\right) = 
\cos^{2N} \left({\Delta_0 \xi \over 2 }\right) = 2^{-2N} \sum_{k = 0}^{2N} C_{2N}^k 
\cos[\Delta_0(N - k) \xi ]
\end{equation}
with  $ C_{2N}^k $ being the binomial coefficients. 
Calculating of integrals gives us the simple formula
\begin{equation}
\langle {\sin \Omega t \over \Omega } \rangle = 2^{-2N} \sum_{m = -N}^{ m = N} C_{2N}^{N-m} 
{\sin(t\sqrt{ 4\Delta_0^2 m^2 + \Delta^2}) \over \sqrt{ 4\Delta_0^2 m^2 + \Delta^2}}.
\end{equation} 
An expression for $\langle (1- \cos \Omega t)/\Omega^2 \rangle $ can be derived by the same approach. 
These formulas should be substituted into Eq.(6) in order to find a time evolution of the qubit's variable 
$\langle \tau_x(t) \rangle .$ 

\section{}

The calculation of the correlation function of $i$-th spin 
$ \langle (1/2) [(\sigma_i)_z(t),(\sigma_i)_z(t_1)]_+\rangle $ in the presence of flipping magnetic field and a heat bath 
is of interest by itself. This problem could be of importance for measuring of the qubit's magnetic field by means of the electronic spin resonance (ESR) in the free radical probe coated on the surface of the superconducting qubit. The flipping of the qubit's magnetic field will be reflected in the characteristics of ESR signal. That is why this method can be used for detecting of coherent and incoherent oscillations in the qubits.

To find the spectral function of $i-th$ nuclear spin we start with the Hamiltonian 
\begin{equation}
H_i = -{\Delta_i\over 2}\tau_x (\sigma_i)_z - (\sigma_i)_x Q_x -  
(\sigma_i)_y Q_y - (\sigma_i)_z Q_z .
\end{equation} 
The spin operators $(\sigma_i)_x,(\sigma_i)_y,(\sigma_i)_z$ obey the following equations 
\begin{eqnarray}
\dot{\sigma_x } = \Delta_i \tau_x \sigma_y - 2 \sigma_z Q_y + 2 \sigma_y Q_z, 
\nonumber\\
\dot{\sigma_y } = -\Delta_i \tau_x \sigma_x + 2 \sigma_z Q_x - 2 \sigma_x Q_z, 
\nonumber\\
\dot{\sigma_z } = - 2 \sigma_y Q_x + 2 \sigma_x Q_y. 
\end{eqnarray}
Here we omit the index $"i"$ in spin operators. The heat bath operators
$(Q_i)_{\alpha }$ are supposed to be independent for different spins $(i = 1,2,..,2N; \alpha =
x,y,z),$ so that the response functions and the correlators of the free heat
bath variables have the form  
\begin{equation}
\langle i [(Q_i)_{\alpha }^{(0)}(t), (Q_j)_{\beta }^{(0)}(t')]_-\rangle \theta
(t-t') =  \varphi (t, t') \delta_{ij} \delta_{\alpha \beta }, 
\end{equation}
\begin{equation}
\langle (1/2) [(Q_i)_{\alpha }^{(0)}(t), (Q_j)_{\beta }^{(0)}(t')]_+ \rangle = 
M(t, t') \delta_{ij} \delta_{\alpha \beta },  
\end{equation}
$\tilde{M}(t,t_1) = M(t,t_1)\theta (t - t_1),$  $\varphi (t, t') = \varphi (t-t'), M(t, t') = M(t-t').$   
As mentioned above, the free heat bath, e.g., phonons without any interaction with spins, is in thermal equilibrium 
at temperature $T.$ In this case the heat bath is characterized by the susceptibility 
$$\chi(\omega ) = \int d\tau e^{i\omega \tau } \varphi (\tau )
$$  
as well as the the spectral density of fluctuations 
$$
S(\omega) = \int d\tau e^{i\omega \tau } M(\tau )
$$
which are related by the fluctuation-dissipation theorem
\begin{equation}
 S(\omega ) = \hbar \chi''(\omega ) \coth(\hbar \omega /2T). 
\end{equation}
The imaginary part of the susceptibility, $\chi''(\omega ), $   describes dissipative properties of the equilibrium heat bath. 

Following the method developed in \cite{16,17,18} we can derive from Eqs.(B2) the Langevin-like
equations for the spin operators, especially, for z-projection $\sigma_z$ we are interested in. 
Due to Gaussian statistics of unperturbed heat bath variables $Q_{\alpha }^{(0)}(t) (\alpha = x,y,z)$ the response of the heat bath to the action of the spin is linear in the spin matrices $\sigma_{\alpha }:$ 
\begin{equation}
Q_{\alpha }(t) = Q_{\alpha }^{(0)}(t) + \int dt_1 \varphi (t-t_1) \sigma_{\alpha }(t_1).
\end{equation}
After substituting these operators into Eqs.(B2) we have to eliminate the unperturbed heat bath variables 
$Q_{\alpha }^{(0)}(t)$ to obtain the equations only for spin variables.   
To do that the products $Q_{\alpha }^{(0)}(t)\sigma_{\beta }(t) (\beta = x,y,z)$ involved in the Heisenberg equations 
(B2) is conveniently represented as 
\begin{equation}
Q_{\alpha }^{(0)}(t)\sigma_{\beta }(t) = \{Q_{\alpha }^{(0)}(t), \sigma_{\beta }(t)\}  + 
\int dt_1 \tilde{M}(t,t_1)i[\sigma_{\beta }(t), \sigma_{\alpha }(t_1)]_-. 
\end{equation}
This formula can be considered as a definition of the brackets $\{..\}$ having zero mean value: 
$$\langle \{Q_{\alpha }^{(0)}(t), \sigma_{\beta }(t)\}\rangle = 0 $$ 
because of the quantum Furutsu-Novikov theorem \cite{16}. Introducing the fluctuation forces, for example, 
\begin{equation}
\xi_z(t) = 2 \{Q_y^{(0)}(t), \sigma_x(t) \} - 2 \{Q_x^{(0)}(t), \sigma_y(t) \}
\end{equation}
with zero mean values, $\langle \xi_z(t) \rangle = 0,$ we obtain the Langevin-like equation for z-projection 
of $i$-th spin:
\begin{eqnarray}
\dot{\sigma_z} = \xi_z - \nonumber\\
 2 \int dt_1 \{\tilde{M}(t,t_1) i [\sigma_y(t), \sigma_x(t_1)]_- +
\varphi(t,t_1) (1/2)[\sigma_y(t), \sigma_x(t_1)]_+ \} + \nonumber\\
2 \int dt_1 \{\tilde{M}(t,t_1) i [\sigma_x(t), \sigma_y(t_1)]_- +
\varphi(t,t_1) (1/2)[\sigma_x(t), \sigma_y(t_1)]_+ \}.
\end{eqnarray}
The brackets $\{..\}$ simplify also the problem of calculating correlations functions of the fluctuation forces. 
They merely imply that  pairings of operators within the same brackets should be omitted in the process. 
As a result, the explicit expression for the fluctuation force $\xi_z $ (B8) together with the definition (B7) allow us to 
get the following correlator    
\begin{equation}
\langle (1/2)[ \xi_z(t),\xi_z(t')]_+\rangle = 2 M(t,t') \langle [\sigma_x(t),
\sigma_x(t')]_+ + [\sigma_y(t),\sigma_y(t')]_+ \rangle .
\end{equation}
It should be emphasized that this result is not based on the fluctuation-dissipation theorem 
and will be valid in the case of strong non-equilibrium as well. 
However, here we resort to the approximation of weak coupling between the spin an the heat bath.
With this approximation we are able to calculate the (anti)commutators in Eqs.(B9),(B10) by means of the equations 
describing the spin evolution in the time-varying qubit's magnetic field:
\begin{eqnarray}
\sigma_x(t) = \sigma_x(t_1) \cos\Lambda_i(t,t_1) + \sigma_y(t_1)
\sin \Lambda_i(t,t_1), \nonumber\\
 \sigma_y(t) = \sigma_y(t_1) \cos\Lambda_i(t,t_1) - \sigma_x(t_1)
\sin \Lambda_i(t,t_1), \nonumber\\
\sigma_z(t) = \sigma_z(t_1),
\end{eqnarray}
where 
\begin{equation}
\Lambda_i(t,t_1) = \Delta_i \int_{t_1}^t dt_2 \tau_x(t_2).
\end{equation}
The operators (B11) obey the equations (B2) where all heat bath variables $Q_{\alpha }$ are omitted.  
Taking into consideration a free oscillations of the qubit operator $\tau_x(t)$ with the tunneling frequency $\Delta$  
we obtain for the function 
$\Lambda_i(t,t_1):$ 
\begin{equation}
\Lambda_i(t,t_1) = {\Delta_i\over \Delta} \left[ \tau_x(t)  \sin\Delta(t-t_1)  + 
 \tau_y(t)  (1 - \cos \Delta(t - t_1)) \right] .
\end{equation}
With the properties of Pauli matrices we find also that 
\begin{equation}
\cos\Lambda_i(t,t_1) = \cos \left( 2 {\Delta_i \over \Delta }
 \sin\left[{\Delta(t-t_1) \over 2}\right] \right),
\end{equation}
\begin{eqnarray}
\sin \Lambda_i(t,t_1) = 2\sin \left( 2 {\Delta_i \over \Delta}\sin\left[{\Delta(t-t_1) \over 2}\right] \right) \times 
\nonumber\\
\left(\tau_x(t) \cos\left[{\Delta (t-t_1) \over 2}\right]   + \tau_y(t)  \sin\left[{\Delta (t-t_1) \over
2}\right] \right).
\end{eqnarray}
As a result, for z-projection of $i$-th spin operator we obtain the simple
stochastic equation:
\begin{equation}
\dot{\sigma}_z(t) + \int dt_1 \gamma_i(t,t_1) \sigma_z(t_1) = \xi_z(t) +
v_i(t).
\end{equation}
Here 
\begin{eqnarray}
\gamma_i(t,t_1) = 8 \tilde{M}(t,t_1) \cos\Lambda_i(t,t_1), \nonumber\\
v_i(t) = 4 \int dt_1 \varphi(t,t_1) \sin \Lambda_i(t,t_1).
\end{eqnarray}
The relaxation of the $i$-th nuclear spin is determined by the real part
$\gamma_i(\omega) $ of the Fourier transform of the function
$\gamma_i(t,t_1):$
\begin{equation}
\gamma_i(\omega) = 4 \int_{-\infty }^{+\infty } d\tau M(\tau ) \cos (\omega
\tau ) \cos \left[ 2{\Delta_i \over \Delta} \sin\left({\Delta \tau \over 2}\right) \right].
\end{equation}
After a short calculation we derive the following formula for the
function $\gamma_i(\omega):$
\begin{equation}
\gamma_i(\omega) = 2 \sum_m J_m\left( 2 {\Delta_i \over \Delta }\right) 
\left[ S\left(\omega - {m\Delta \over 2} \right) + S\left(\omega + {m\Delta
\over 2} \right)\right].
\end{equation} 
Here $J_m(z)$ is the Bessel function of m-th order, and $ S(\omega ) $ is a 
spectral density of the heat bath fluctuations (B5).
It should be emphasized that it is the formula (B19) that gives a line width of the spin-1/2 coupled to a heat bath 
in the presence of the flipping magnetic field created by the qubit. This line width can be measured in 
experiments on electronic or nuclear paramagnetic resonance. 

It follows from Eqs.(B10),(B11) that the Fourier transform of the correlation function of
the fluctuation forces is also determined by the formula like Eq.(B19):
\begin{equation}
\langle {1\over 2}[\xi_z^{(i)}(\omega ), \xi_z^{(i)}]_+\rangle = 2
\gamma_i(\omega).
\end{equation}
As a result 
the spectral density of fluctuations of $i-th$ nuclear spin 
$$
K_i(\omega ) = \int dt \langle {1\over 2} [(\sigma_i)_z(t),(\sigma_i)_z(0)]_+\rangle e^{i\omega t} 
$$
is determined by the expression
$$
K_i(\omega ) = {2\gamma_i(\omega ) \over \omega^2
+ \gamma_i^2(\omega ) }, 
$$
whereas
the spectral density (15) has the simple form:
\begin{equation}
S_P(\omega ) = {1\over 2 } \sum_i \Delta_i^2 {\gamma_i(\omega ) \over \omega^2
+ \gamma_i^2(\omega ) }
\end{equation}
with the function $\gamma_i(\omega)$ given by Eq.(B19).

\end{document}